\documentclass[onecolumn,preprintnumbers,12pt,nofootinbib,showpacs]{revtex4}

\usepackage{graphicx}% Include figure files
\usepackage{bm}% bold math
\usepackage{dsfont}
\usepackage{mathrsfs} % font for mathscr-command (rsfs: Raph Smith’s For­mal Script)
\usepackage{amssymb,amsfonts,amsmath}
\usepackage{graphicx}
\usepackage{color}
\newcommand{\dd}{\mathrm{d}}
\newcommand{\ee}{\mathrm{e}}

\newcommand{\R}{\mathds R}

\newcommand{\Scri}{\mathscr{I}}

\normalsize

\begin{document}

\title{\Large Fully pseudospectral solution of the conformally invariant wave equation near the cylinder at spacelike infinity}

\author{\bf J\"org Frauendiener}
\email{joergf@maths.otago.ac.nz}
\author{\bf J\"org Hennig}
\email{jhennig@maths.otago.ac.nz}

\affiliation{Department of Mathematics and Statistics,
           University of Otago,
           P.O. Box 56, Dunedin 9054, New Zealand}

\begin{abstract}
 We study the scalar, conformally invariant wave equation on a four-dimensional Minkowski background in spherical symmetry, using a fully pseudospectral numerical scheme. Thereby, our main interest is in a suitable treatment of spatial infinity, which is represented as a cylinder. We consider both Cauchy problems, where we evolve data from a Cauchy surface to future null infinity, as well as characteristic initial value problems with data at past null infinity, and demonstrate that highly accurate numerical solutions can be obtained for a small number of grid points.
% \\[2ex] [manuscript date: {\color{red} \today, \currenttime}]
\end{abstract}

%04.20.Ha 	GR: Asymptotic structure 
%04.25.D- 	GR: Numerical relativity
%02.70.Jn 	Computational techniques; simulations: Collocation methods 

\maketitle
%%%%%%%%%%%%%%%%%%%%%%%%%%%%%%%%%%%%%%%%%%%%%%%%%%%%%%%%%%%%%%%%%%%%%%%%%%%%%
\section{Introduction\label{sec:intro}}

The \emph{global} treatment of asymptotically flat (or, more generally, asymptotically simple) spacetimes as models for isolated systems is an important topic, both in mathematical and in numerical relativity. In particular, this is relevant for a rigorous investigation of gravitational wave signals, which can be uniquely read off only at null infinity $\Scri$. Indeed, it has been shown that certain tail decay rates at timelike boundaries (which are often introduced as ``artificial'' outer boundaries for numerical computations) and at null infinity are different, so that an appropriate treatment of infinity is crucial, see, e.g., \cite{Gundlach1994, Zenginoglu2008} and references therein.

An important ingredient for such a rigorous treatment is Penrose's conformal boundary \cite{Penrose1964,Penrose1965}, which is obtained by isometrically embedding the spacetimes into larger Lorentzian manifolds and conformally rescaling the metric, i.e.\ the ``physical'' metric is replaced by a conformal metric and a conformal factor. A formulation of Einstein's field equations adapted to this setting, the conformal field equations (CFE), was given by Friedrich \cite{Friedrich1983,Friedrich1986}. In the context of this approach, it is well-known that certain mathematical difficulties arise in the region near spacelike infinity $i^0$. In order to cope with this problem, Friedrich has introduced yet another formulation of the field equations, the generalised CFE \cite{Friedrich1998}. Moreover, a novel representation of spacelike infinity has been given: $i^0$, in regular cases thought of before as a point, is ``blown up'' to a cylinder $S^2\times[-1,1]$, which connects past and future null infinity $\Scri^\pm$.

The cylinder-formulation is very appealing for numerical calculations as well. Recently, the ``core'' equations of the generalized CFE, a spin-2 system on Minkowski background describing linearised gravitational fields near spacelike infinity, have been solved numerically with a finite-difference method (in particular, using fourth-order Runge-Kutta in time), see \cite{BeyerDoulis2012,BeyerDoulis2013,Doulis2013}.
A typical drawback in this approach (and other finite-difference schemes) is the restriction of the time step by the Courant-Friedrichs-Lewy (CFL) condition. In particular, as one approaches $\Scri$, the characteristic speeds of the system diverge. Consequently, the requirement that the numerical domain of dependence contain the analytical one enforces smaller and smaller time steps. Hence, it is only possible to come close to $\Scri$, using a very large number of grid points in the time direction, without actually reaching it. However, since gravitational wave signals are defined at $\Scri$ itself, it is desirable to include this boundary into the numerical domain.

In the present article, we address this problem in a simplified setting: we consider the conformally invariant wave equation on a Minkowski background. Moreover, we restrict ourselves to the $1+1$ dimensional problem in spherical symmetry. The resulting wave equations are simpler than the above mentioned spin-2 system, but mirror already important mathematical properties and difficulties of the general problem. Hence, a successful numerical treatment of this model will lead the way to future applications to the linearised, or even to the full Einstein equations.

Our main objective is to demonstrate that the wave equation near the cylinder at spacelike infinity (and even in the entire Minkowski space) can be solved with a fully pseudospectral scheme.\footnote{For comparison with numerical investigations of conformally invariant wave equations using a \emph{finite-difference} approach, we refer the reader to \cite{CruzOsorio2010}.} 
The underlying numerical method, which is based on the approach by Ansorg \emph{et~al.} \cite{Ansorg2002,Ansorg2003} for solving elliptic PDEs, was introduced in \cite{Hennig2009} and later applied to several physical problems in \cite{Ansorg2011,Hennig2013}. The main idea is to expand the unknown functions in terms of Chebyshev polynomials, both with respect to space \emph{and} time. This provides a highly implicit method for which, in particular, no CFL condition is required, since the solution is simultaneously obtained in the entire numerical domain. Consequently, we will see that we can reach $\Scri^+$ with a small number of grid points (about 20-30 points in time direction) instead of only approaching it asymptotically.

In the following, we solve \emph{Cauchy problems} with data on a hypersurface $t=0$ and evolve them to future null infinity $\Scri^+$. Thereby, we first concentrate on a vicinity of $i^0$ only, and afterwards study the same problem for all values of a radial coordinate $r$, i.e.\ in the entire Minkowski space and not only close to infinity. 

A different type of initial value problem that has frequently been studied in general relativity and that is again gaining more attention in recent times, is an \emph{(asymptotic) characteristic initial value problem with data on past null infinity} $\Scri^-$. We refer to \cite{Friedrich1981,Friedrich1986a,Kannar1996,Friedrich2013,Chrusciel2013} 
for some interesting mathematical results. Here, we demonstrate that this type of initial value problem, applied to the conformally invariant wave equation, can also be solved with the fully pseudospectral method. The different initial value problems considered here are illustrated in Fig.~\ref{fig:IVPs} below.

This paper is organized as follows. We begin with a quick overview of the conformally invariant wave equation and the fully pseudospectral scheme in Sec.~\ref{sec:background}. Afterwards, in Sec.~\ref{sec:i0}, we apply this method to the wave equation near spacelike infinity. In Sec.~\ref{sec:all}, we consider different coordinates that cover all of Minkowski space and again solve the wave equation. Finally, in Sec.~\ref{sec:discuss}, we summarize our results.

%%%%%%%%%%%%%%%%%%%%%%%%%%%%%%%%%%%%%%%%%%%%%%%%%%%%%%%%%%%%%%%%%%%%%%%%%%%%%
\section{Background\label{sec:background}}

%%%%%%%%%%%%%%%%%%%%%%%%%%%%%%%%%%%%%%%%%%%%%%%%%%%
\subsection{The conformally invariant wave equation\label{sec:ConfWE}}

In $n$ spacetime dimensions, the conformally invariant wave equation is
\begin{equation}
 g^{ab}\nabla_a\nabla_b f-\frac{n-2}{4(n-1)}R f=0,
\end{equation}
where $g^{ab}$ is the (inverse) metric and $R$ the corresponding Ricci scalar. This equation is conformally invariant for $f$ with conformal weight $s=1-n/2$, i.e.\ if $\tilde f$ solves the equation for a metric $\tilde g$, then 
$f=\Theta^s\tilde f$ solves the equation for a conformally related metric $g_{ab}=\Theta^2 \tilde g_{ab}$ with conformal factor $\Theta$. 

Here we consider the equation in Minkowski space, i.e.\ we assume that the physical metric $\tilde g_{ab}=\eta_{ab}$ is the Minkowski metric, and we restrict ourselves to $n=4$ dimensions, where we have the equation
\begin{equation}\label{eq:ConfWE}
 g^{ab}\nabla_a\nabla_b f-\frac{R}{6}f=0.
\end{equation}
Then the conformal weight is $s=-1$. Moreover, we consider only spherically symmetric solutions $f=f(t,r)$, which depend on time $t$ and on a radial coordinate $r$ alone. 

In this article we look at Cauchy problems, where the function values and first time derivatives of $f$ can be prescribed at the initial Cauchy surface. We also study characteristic initial value problems, where only the initial function values can be specified.

%%%%%%%%%%%%%%%%%%%%%%%%%%%%%%%%%%%%%%%%%%%%%%%%%%%
\subsection{Fully pseudospectral time evolution\label{sec:spectral}}

In the following we give a short summary of the fully pseudospectral scheme, which will enable us to find highly accurate numerical solutions to the conformally invariant wave equation. For more details we refer to \cite{Hennig2009}. 

The basic idea is to express the unknown function(s) in terms of truncated series of Chebyshev polynomials $T_k$, both with respect to space and time, and to reduce the differential equation (in our case the conformally invariant wave equation) to an algebraic system of equations. 

To this end, the physical domain is first mapped to a unit square by introducing spectral coordinates $(\sigma,\tau)\in[0,1]\times[0,1]$ such that the initial surface corresponds to $\tau=0$. Then we enforce the initial conditions by expressing the unknown function $f(\sigma,\tau)$ in terms of another unknown $f_2(\sigma,\tau)$ with an expansion
\begin{equation}\label{eq:expan}
 f(\sigma,\tau)=\begin{cases}
                 f_0(\sigma)+\tau f_1(\sigma)+\tau^2 f_2(\sigma,\tau),
                  & \textrm{if function values and first derivatives are given}\\
                 f_0(\sigma)+\tau f_2(\sigma,\tau),
                  & \textrm{if only function values are given}
                \end{cases}
\end{equation}
depending on the type of initial data (Cauchy problem or characteristic initial value problem). The functions $f_0(\sigma)$ and $f_1(\sigma)$ can be obtained from the given initial data  at $\tau=0$.

In a next step, we approximate the new unknown $f_2$ via
 \begin{equation}\label{eq:approx}
   f_2(\sigma,\tau)\approx\sum\limits_{j=0}^{N_\sigma}\sum_{k=0}^{N_\tau}
        c_{jk}T_j(2\sigma-1)T_k(2\tau-1)
\end{equation}
for given spectral resolution (number of polynomials) $n_\sigma \equiv N_\sigma+1$ and $n_\tau \equiv N_\tau+1$. From this approximation for $f_2$ we can then easily find approximations for $f$ and its derivatives. 
 
An algebraic system of equations for the Chebyshev coefficients $c_{jk}$ can be obtained by requiring that the wave equation (and suitable boundary or regularity conditions, if applicable) are satisfied at a set of collocation points. Here, we choose \emph{Gauss-Lobatto} nodes, which are defined by
\begin{equation}\label{eq:nodes}
 \sigma_j=\sin^2\left(\frac{\pi j}{2N_\sigma}\right),\quad
 \tau_k=\sin^2\left(\frac{\pi k}{2N_\tau}\right),\quad
 j=0,\dots,N_\sigma,\quad
 k=0,\dots,N_\tau.
\end{equation}
Note that, instead of considering the resulting algebraic system of equations as a system for $c_{jk}$, we actually solve directly for the function values 
$f_{jk}\equiv f_2(\sigma_j,\tau_k)$ at the nodes. 

Finally, starting from some initial guess (in our examples we can choose the trivial solution $f_{jk}=0$ for all $j,k$), we solve this system iteratively with the \emph{Newton-Raphson} method. 

%%%%%%%%%%%%%%%%%%%%%%%%%%%%%%%%%%%%%%%%%%%%%%%%%%%%%%%%%%%%%%%%%%%%%%%%%%%%%
\section{The wave equation near spacelike infinity\label{sec:i0}}

%%%%%%%%%%%%%%%%%%%%%%%%%%%%%%%%%%%%%%%%%%%%%%%%%%%
\subsection{Coordinates, wave equation and boundary conditions\label{sec:i0coord}}

In order to describe a vicinity of spacelike infinity $i^0$ and to resolve the cylinder-like structure of $i^0$, we first introduce suitable coordinates. To this end, we start from the physical metric $\tilde g$ of Minkowski space, which takes the form
\begin{equation}\label{eq:metric1}
 \tilde g = \dd\tilde t^{\,2}-\dd\tilde r^2
            -\tilde r^2(\dd\theta^2+\sin^2\!\theta\,\dd\phi^2)
\end{equation}
in spherical coordinates $(\tilde t, \tilde r, \theta,\phi)$. Then we define new coordinates $t$ and $r$ as
\begin{equation}
 t = \frac{\tilde t}{\tilde r},\quad
 r = -\frac{\tilde r}{\tilde t^2-\tilde r^2}.
\end{equation}
The metric can now be expressed as $\tilde g = \Theta^{-2} g$ in terms of the conformal metric
\begin{equation}\label{eq:metric2}
 g=\dd t^2+2\frac{t}{r} \dd t\dd r - \frac{1-t^2}{r^2}\dd r^2 - \dd\sigma^2
\end{equation}
with the conformal factor
\begin{equation}
 \Theta=r(1-t^2).
\end{equation}
Note that this is a special case of the more general coordinates considered by Friedrich \cite{Friedrich2003}, corresponding to the choice $\mu(r)\equiv 1$ for the free function $\mu(r)$ that appears in Friedrich's transformation. 

\begin{figure}\centering
 \includegraphics[width=\linewidth]{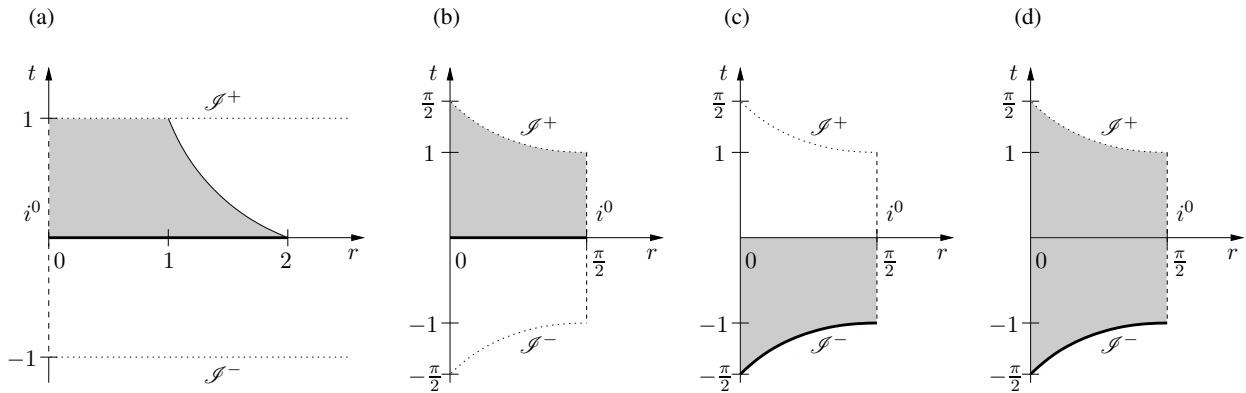}
 \caption{\label{fig:IVPs}
 Illustration of the two types of Cauchy problems (a, b) and the two types of characteristic initial value problems (c, d) under consideration. Initial data are prescribed at the thick lines and evolved into the shaded regions. Spatial infinity $i^0$ is indicated with dashed lines, and dotted lines show the positions of null infinity $\Scri^+$ and $\Scri^-$. The coordinates in (a) are adopted for studying a vicinity of $i^0$ (see Sec.~\ref{sec:i0}), whereas the coordinates in (b-d) cover the entire Minkowski space (see Sec.~\ref{sec:all}).}
\end{figure}

Now we can look at the conformally invariant wave equation corresponding to the metric \eqref{eq:metric2}. Since the Ricci scalar $R$ of this metric vanishes identically, we only have to consider the first term in \eqref{eq:ConfWE}. This leads to
\begin{equation}\label{eq:WE1}
 0 = (1-t^2) f_{,tt}+2tr f_{,tr}-r^2f_{,rr}-2tf_{,t},
\end{equation}
where a comma denotes partial derivatives.

In our setting, the coordinates are defined in the domain $t\in[-1,1]$, $r\ge0$. The boundary $t=\pm1$, $r>0$ corresponds to null infinity $\Scri^\pm$, and $i^0$ is located at $r=0$, see Fig.~\ref{fig:IVPs}a.

The characteristics of the wave equation \eqref{eq:WE1} are lightlike curves with respect to the metric $g$ and have the form
\begin{equation}
 t=\frac{c}{r}\pm 1, \quad c\in\R,
\end{equation}
where different values of the constant $c$ correspond to different characteristics.
We intend to solve a Cauchy problem with data at $t=0$, which will then be evolved until $t=1$, and we would like to obtain the solution $f$ in some vicinity of $i^0$, say for $0\le r\le 1$. On the other hand, if we strictly restrict ourselves to the domain $(t,r)\in[0,1]\times[0,1]$, then we would need to prescribe boundary conditions at the outer boundary $r=1$. However, since there is no physical reason why $f$ should satisfy any particular conditions at $r=1$, ``artificial'' conditions at this boundary --- even though mathematically perfectly valid --- would be rather unphysical. Therefore, we instead choose a slightly larger domain that contains the square $[0,1]\times[0,1]$, but where the outer boundary is the ingoing characteristic $t=2/r-1$, see the shaded region in Fig.~\ref{fig:IVPs}a. Then there is no need for boundary conditions at this curve. Instead, we can impose the wave equation itself there.

At $r=0$ (i.e.\ at $i^0$) the wave equation \eqref{eq:WE1} becomes an \emph{intrinsic} equation,
\begin{equation}
 0 = (1-t^2)f_{,tt}-2t f_{,t},
\end{equation}
in accordance with the observation that the cylinder at spacelike infinity is a \emph{totally characteristic surface} also for Friedrich's conformal field equations.
The latter equation has the first integral $f_{,t}=d/(1-t^2)$, $d\in\R$. This shows that $f$ will be singular at $r=0$, $t=1$, i.e.\ the point where $i^0$ and $\Scri^+$ intersect\footnote{A degeneracy at the intersection between the cylinder at spatial infinity and null infinity, usually denoted as $I^\pm$, also is a well-known property of Friedrich's conformal field equations, see \cite{Friedrich2003}. But as we see, this problem already shows up in the wave equation discussed here. In both cases, the requirement of regularity leads to regularity conditions in the form of restrictions of the initial data.}, unless we choose initial data at $t=0$ with $f_{,t}=0$ at $r=0$ (for which $d=0$). In the following, we restrict ourselves to this choice, in which case $f$ will be constant at $i^0$. Therefore, for our numerical method, we can choose $f_{,t}=0$ as a regularity condition there.

Another condition at $i^0$ follows from first differentiating \eqref{eq:WE1} with respect to $r$ and then performing the limit $r\to0$. This leads to $f_{,ttr}=0$, i.e.\ $f_{,tr}=\textrm{constant}$ at $r=0$. We will use this condition (instead of the condition $f_{,t}=0$) at the single point\footnote{If we restrict the numerical evolution to a time interval $[0,t_\textrm{max}]$ with some $t_\textrm{max}<1$, then the condition $f_{,t}=0$ is sufficient at $t=t_\textrm{max}$, $r=0$. For $t_\textrm{max}=1$, however, it turns out that the alternative condition \eqref{eq:cond1} needs to be imposed in order to guarantee numerical stability.} $t=1$, $r=0$ in the form
\begin{equation}\label{eq:cond1}
 f_{,tr}(1,0)=f_{,tr}(0,0),
\end{equation}
where the right hand side is obtained from the given values for $f_{,t}$ at $t=0$.

We can easily derive the general solution to the wave equation \eqref{eq:WE1}, which is useful for testing the accuracy of our numerical method. Starting from the solution 
$\tilde f(\tilde t,\tilde r) = [\tilde F(\tilde r+\tilde t)+\tilde G(\tilde r-\tilde t)]/\tilde r$ 
of the wave equation with respect to the physical metric $\tilde g$, where $\tilde F$ and $\tilde G$ are two free functions\footnote{If we require regularity of the solution $\tilde f$ at the origin $\tilde r=0$ of Minkowski space, then we need to choose $\tilde G(x)=-\tilde F(-x)$. However, since we are currently only interested in a vicinity of spatial infinity, we can choose $\tilde F$ and $\tilde G$ independently.}
we obtain (cf.~Sec.~\ref{sec:ConfWE})
\begin{equation}\label{eq:gensol1}
 f(t,r)=\Theta^{-1}\tilde f(\tilde t,\tilde r)
       =A\big(r(1-t)\big)+B\big(r(1+t)\big),
\end{equation}
in terms of free functions $A$ and $B$ (related to $\tilde F$ and $\tilde G$ via $A(x)=\tilde F(1/x)$ and $B(x)=\tilde G(1/x)$).

%%%%%%%%%%%%%%%%%%%%%%%%%%%%%%%%%%%%%%%%%%%%%%%%%%%
\subsection{Numerical solution}
 
We apply the fully pseudospectral scheme as described in Sec.~\ref{sec:spectral}. For that purpose, we introduce spectral coordinates $\sigma,\tau\in[0,1]$ via
\begin{equation}
 t=\tau,\quad r=\frac{2\sigma}{1+\tau}.
\end{equation}
The outer boundary (the ingoing characteristic) is then given by $\sigma=1$, whereas $\sigma=0$ corresponds to $i^0$.

As initial data we prescribe the values of $f$ and $f_{,t}$ at $t=\tau=0$, which is numerically implemented with an expansion of the form \eqref{eq:expan}. As discussed above, we impose the boundary condition $f_{,t}=0$ at $r=\sigma=0$, with exception of the point $\sigma=0$, $\tau=1$, where we use condition \eqref{eq:cond1}. Another exceptional point is $\sigma=\tau=0$ on the initial surface, where $f_{,t}=0$ is automatically satisfied, because, as discussed above, we choose initial data subject to this condition. At this point, we instead use the condition $f_{,tt}=0$ (which is a simple consequence of $f_{,t}=0$ at $r=0$). At all other grid points we impose the conformally invariant wave equation \eqref{eq:WE1}.

In order to test the numerical accuracy of our method, we choose three different exact solutions by specifying the free functions $A$ and $B$ in the general solution \eqref{eq:gensol1} as given in Table~\ref{tab:ExA}.

\begin{table}[ht]
 \begin{tabular}{|p{2.5cm}p{2.5cm}p{2.5cm}|}
  \hline
            & $A(x)$              & $B(x)$\\
  \hline
  Example 1: & $\sin(3x)$          & $\sin(3x)$\\[1ex]
  Example 2: & $\ee^{-x}$          & $1$\\[1ex]
  Example 3: & $\ee^{-2(x-1)^2}+x$ & $\ee^{-2(x-1)^2}-x$\\
  \hline  
 \end{tabular}
 \caption{\label{tab:ExA}
  The functions $A(x)$ and $B(x)$ for three example solutions to the conformally invariant wave equation \eqref{eq:WE1} near spatial infinity, cf.~Eq.~\eqref{eq:gensol1}.}
\end{table}

For each of the three examples we compute the initial data at $t=0$ from \eqref{eq:gensol1}, solve the problem numerically and then determine the numerical error $|f_\textrm{numerical}-f_\textrm{exact}|$ at $100\times 100$ equidistant grid points via Chebyshev interpolation, i.e.\ we compare the solutions not only at the collocation points but at general points in between. The maximum value of the errors at these $10{,}000$ points, which estimates the $L^\infty$-norm of $f_\textrm{numerical}-f_\textrm{exact}$, is then used as a measure for the overall accuracy of the method.

\begin{figure}\centering
 \includegraphics[width=\linewidth]{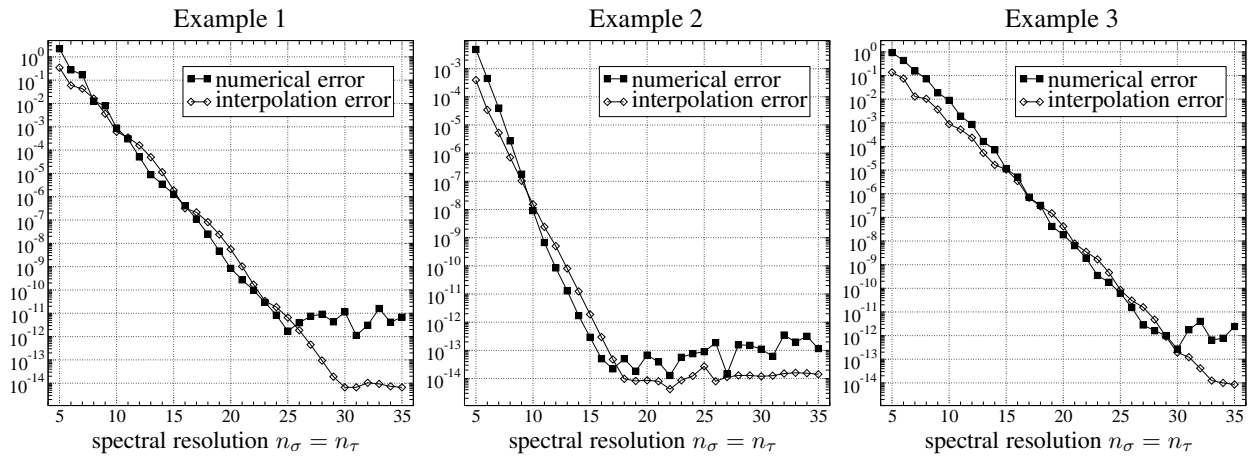}
 \caption{\label{fig:ConA}
  Convergence plots for the Cauchy problem near spacelike infinity, which is illustrated in Fig.~\ref{fig:IVPs}a. The numerical error and, for comparison, the general error of Chebyshev interpolation are shown for the three numerical examples from Table~\ref{tab:ExA}.}
\end{figure}

In all examples, we solve the problem with different numerical resolutions $n_\sigma$, $n_\tau$ (numbers of collocation points, cf.~\eqref{eq:approx}, \eqref{eq:nodes}) and investigate how the error changes. As an indication for the maximum accuracy that could be expected in our method (or rather as an upper bound for the accuracy), we also compute the error merely due to Chebyshev interpolation of the given exact solution. To this end we read off the exact function values at the collocation points, compute the values at the $100\times100$ grid points via Chebyshev interpolation (without solving any differential equations) and compare with the exact values at these points. The corresponding convergence plots are shown in Fig.~\ref{fig:ConA}.

We observe the typical features as expected for this method: the curves fall off approximately linearly in a logarithmic scale,  indicating exponential convergence, until a final saturation level close to machine accuracy (of our double-precision code) is reached. The final numerical error is about 1-3 orders of magnitude larger than the pure interpolation error, since the additional error of the numerical method is involved\footnote{Interestingly, we observe that for some particular resolutions, the numerical error is even slightly smaller than the interpolation error. This comes from the fact that the Chebyshev interpolation polynomial is not identical with, but only close to the \emph{optimal} interpolation polynomial of a given function. Hence it can happen that our numerical method, which produces a linear combination of Chebyshev polynomials, leads to a polynomial that is closer to the optimal polynomial then the Chebyshev interpolation polynomial. But this is not expected in general.}.

\begin{figure}\centering
 \includegraphics[width=0.6\linewidth]{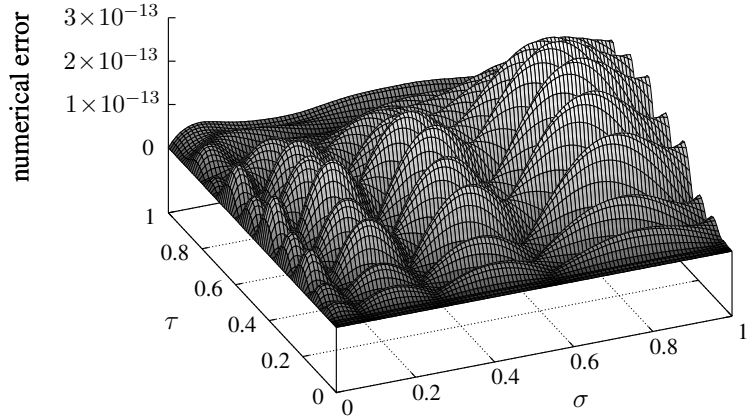}
 \caption{\label{fig:ErrA}
 The numerical error for the wave equation near $i^0$ is shown as a function of the spectral coordinates $\sigma$ and $\tau$. Here we have considered the example solution 2 from Table~\ref{tab:ExA} with resolution $15\times 15$.}
\end{figure}

It is also interesting to see how the errors are distributed over the numerical domain and to identify the regions with the largest error. To this end, for a given example and fixed spectral resolution, we plot the error $|f_\textrm{numerical}-f_\textrm{exact}|$ as a function of the spectral coordinates $\sigma$ and $\tau$. One such plot is shown in Fig.~\ref{fig:ErrA}, and plots for other resolutions or other initial data look quite similar and show the same qualitative behaviour.
We see that the error is zero at $t=0$, which is clear since we enforce the initial conditions exactly with the expansion \eqref{eq:expan}. 
In a conventional numerical time-stepping scheme, where the initial data are evolved step-by-step from time-slice to time-slice, one generally observes an overall upward trend of the numerical error with increasing time. In our highly implicit scheme, where in principle all collocation points are coupled, this is not the case.
Indeed, we see that the largest error is not assumed at $\tau=1$, but at or near the outer boundary $\sigma=1$ (depending on the particular example; for the plot in Fig.~\ref{fig:ErrA} the maximum error is found at $\sigma=0.80$, $\tau = 0.45$).

%%%%%%%%%%%%%%%%%%%%%%%%%%%%%%%%%%%%%%%%%%%%%%%%%%%%%%%%%%%%%%%%%%%%%%%%%%%%%
\section{Entire Minkowski space\label{sec:all}}

So far, since we have focused on a vicinity of spatial infinity $i^0$, we have considered coordinates that cover only a part of Minkowski space. Now we introduce different coordinates that allow us to study the wave equation in all of Minkowski space, but still ``blow up'' $i^0$ to a cylinder, which is important for a rigorous treatment of spatial infinity.

%%%%%%%%%%%%%%%%%%%%%%%%%%%%%%%%%%%%%%%%%%%%%%%%%%%
\subsection{Coordinates, wave equation and boundary conditions}

Again we start from the Minkowski metric $\tilde g$ in spherical coordinates, see Eq.~\eqref{eq:metric1}. Then we perform the usual conformal compactification that leads to the Penrose diagram where (suppressing the angles $\theta$ and $\phi$) Minkowski space is represented by a triangle, see, e.g., the overview article \cite{Frauendiener2004}. To this end, we first introduce null coordinates 
$\tilde u=\tilde t-\tilde r$, $\tilde v=\tilde t+\tilde r$, compactify them via
$\tilde u=\tan u$, $\tilde v=\tan v$ and finally set
$T=v+u$, $R=v-u$. The coordinates $(T,R)$ obtained in this way cover the entire Minkowski space, but $i^0$ is represented as the single point $T=0$, $R=\pi$. In order to resolve the internal structure of $i^0$, we perform an additional coordinate transformation
\begin{equation}
 T = 2t\kappa(r),\quad
 R = 2r\quad\textrm{with}\quad \kappa(r)=\cos(r)
\end{equation}
thus introducing new coordinates $(t,r)$, which are defined in the intervals 
\begin{equation}\label{eq:domain}
 0\le r\le \frac{\pi}{2},\quad 
 -t_\textrm{max}(r)\le t\le t_\textrm{max}(r),\quad\textrm{where}\quad
 t_\textrm{max}(r):=\frac{\frac{\pi}{2}-r}{\cos r}.
\end{equation}
(We could make other choices for the function $\kappa(r)$ in order to change the coordinate location of $\Scri^\pm$, but here we restrict ourselves to the above case.) The Minkowski metric now takes the form $\tilde g=\Theta^{-2} g$ with
\begin{equation}\label{eq:metric3}
 g = \dd t^2-2t\tan r\,\dd t\dd r-[1+(1-t^2)\tan^2\!r]\dd r^2-\sin^2\!r(\dd\theta^2+\sin^2\!\theta\,\dd\phi^2)
\end{equation}
and
\begin{equation}
 \Theta = \frac{\cos(t\kappa-r)\cos(t\kappa+r)}{\cos r}.
\end{equation}
In these coordinates, the origin of Minkowski space is given by $r=0$, $i^0$ is located at $r=\pi/2$ and $\Scri^\pm$ corresponds to $t=\pm t_\textrm{max}(r)$, see Fig.~\ref{fig:IVPs}b-d. 

The Ricci scalar of the metric \eqref{eq:metric3} turns out to be $R=-6\cos^2\!r$ and the conformally invariant wave equation \eqref{eq:ConfWE} becomes
\begin{equation}\label{eq:WE2}
%  \begin{aligned}
  (1-t^2\sin^2\!r)f_{,tt}-2t\sin r\cos rf_{,tr}-\cos^2\!r\, f_{,rr}
   -t(2+\cos^2\!r)f_{,t}-2\frac{\cos^3\! r}{\sin r}f_{,r}
  +\cos^2\!r\, f = 0.
%  \end{aligned}
\end{equation}

The general solution can again be derived from the solution $\tilde f$ to the wave equation with respect to the physical metric $\tilde g$. We obtain
\begin{equation}
 f(t,r) = \Theta^{-1}\tilde f(\tilde t,\tilde r)
        = \frac{F(t\cos r-r)+G(t\cos r+r)}{\sin(r)}
\end{equation}
in terms of two free functions $F$ and $G$. The condition of regularity at $r=0$ leads to the restriction $G(x)=-F(x)$ and we finally have
\begin{equation}\label{eq:gensol2}
 f(t,r) = \frac{F(t\cos r-r)-F(t\cos r+r)}{\sin(r)}.
\end{equation}

Multiplying the wave equation \eqref{eq:WE2} with $\sin r$ and then performing the limit $r\to0$ leads to the regularity condition
\begin{equation}
 f_{,r}=0\quad\textrm{at}\quad r=0.
\end{equation}
Furthermore, similar to the discussion in Sec.~\ref{sec:i0coord}, the equation at $i^0$, together with the requirement of regularity, results in the boundary condition
\begin{equation}
 f_{,t}=0\quad\textrm{at}\quad i^0.
\end{equation}

In the following we intend to solve three different types of problems:
\begin{enumerate}
 \item[(i)] a \emph{Cauchy problem} with data at $t=0$, which are evolved to $\Scri^+$ (i.e.\ we consider the ``upper half'' of Minkowski space), see Fig.~\ref{fig:IVPs}b,
 \item[(ii)] a \emph{characteristic initial value problem} with data at $\Scri^-$, which are evolved until $t=0$ (``lower half'' of Minkowski space), see Fig.~\ref{fig:IVPs}c, and
 \item[(iii)] a \emph{characteristic initial value problem} with data at $\Scri^-$, which are evolved to $\Scri^+$ (entire Minkowski space), see Fig.~\ref{fig:IVPs}d.
\end{enumerate}

In case of the Cauchy problem (i), the condition $f_{,r}=0$ at $r=0$ will be automatically satisfied at the point $r=t=0$ due to the choice of initial data. A different condition, however, can be derived from the wave equation \eqref{eq:WE2} in the limit $r\to 0$,
\begin{equation}\label{eq:con2}
 f_{,tt}-3f_{,rr}+f=0\quad\textrm{at}\quad r=t=0
\end{equation}
(see \cite{Hennig2009} for further examples of wave equations with ``exceptional points'' at which additional conditions are required). A similar condition can be derived for the characteristic initial value problem with data at $\Scri^-$,
\begin{equation}\label{eq:con3}
 f_{,tt}-3f_{,rr}+\frac{3\pi}{2}f_{,t}+f=0
 \quad\textrm{at}\quad r=0,\ t=-\frac{\pi}{2}.
\end{equation}

Finally, we again find that
\begin{equation}
 f_{,tr}=\textrm{constant}\quad\textrm{at}\quad i^0.
\end{equation}

It is interesting to compare the numerical investigations of the ``regular'' wave equation \eqref{eq:WE2} (which is obtained for the conformal metric $g$) to the \emph{singular} equation corresponding to the physical metric $\tilde g$ (but still expressed in terms of our compactified coordinates $t$ and~$r$), because this allows a study of whether singular terms in a PDE, which certainly hinder analytical considerations, also have negative influences on the numerical results. This equation is
\begin{equation}\label{eq:WE3}
 \begin{aligned}
  0 & = (1-t^2\sin^2\!r)\tilde f_{,tt}-2t\sin r\cos r \tilde f_{,tr}
        -\cos^2\!r\,\tilde f_{,rr}\\
    & \quad 
       -\left[2\cos r\frac{2t\sin^2\!r\,\cos r-\sin(2t\cos r)}
                          {\cos(t\cos r-r)\cos(t\cos r+r)}+3t\cos^2\! r\right]\tilde f_{,t}\\
    & \quad
       -2\cos r\left[\frac{2\sin r\cos^2\! r}{\cos(t\cos r-r)\cos(t\cos r+r)}
       -2\sin r+\frac{1}{\sin r}\right]\tilde f_{,r}.   
 \end{aligned}
\end{equation}
The latter equation has the same principal part as Eq.~\eqref{eq:WE2}, but the coefficients of the first-order derivatives are now singular at $\Scri$, and there is no term proportional to $\tilde f$ since the Ricci scalar of $\tilde g$ vanishes. The general solution, obtained from \eqref{eq:gensol2} via a conformal transformation, is
\begin{equation}\label{eq:gensol3}
 \tilde f(t,r) = \frac{\cos(t\cos r-r)\cos(t\cos r+r)}{\sin r\cos r}
   \left[F(t\cos r-r)-F(t\cos r+r)\right].
\end{equation}
Evaluating \eqref{eq:WE3} at the boundaries, we find the conditions
\begin{equation}
 \tilde f_{,r}=0 \quad\textrm{at}\quad r=0
\end{equation}
and
\begin{equation}\label{eq:con4}
 \tilde f=\textrm{constant}\quad\textrm{at}\quad \Scri^\pm,\quad
 \tilde f_{,t}=\textrm{constant}\quad\textrm{at}\quad i^0.
\end{equation}
We only consider solutions $\tilde f$ that correspond to regular conformally related solutions $f$, which requires that the $\tilde f$ vanishes at zeros of the conformal factor $\Theta$. Hence we have to choose the constants in \eqref{eq:con4} as zero and finally obtain the boundary condition
\begin{equation}
 \tilde f=0\quad\textrm{at}\quad \Scri^\pm, i^0.
\end{equation}
In particular, this fixes the function values of $\tilde f$ at the initial surface $\Scri^-$, which are therefore not available as free initial data. Instead, we must give the values of $\tilde f_{,t}$ there (subject to the mild restriction that $\tilde f_{,t}$ has to vanish at the intersection points of $\Scri^-$ with $r=0$ and with $i^0$, which follows from the wave equation). This is certainly different from typical characteristic initial value problems and comes from the singular nature of the wave equation at $\Scri^-$.

%%%%%%%%%%%%%%%%%%%%%%%%%%%%%%%%%%%%%%%%%%%%%%%%%%%
\subsection{Numerical solution}

In order to solve the Cauchy problem (problem (i) from the previous subsection) and the two types of characteristic initial value problems (problems (ii) and (iii)), we introduce spectral coordinates $\sigma,t\in[0,1]$ as follows 
[cf.\ \eqref{eq:domain}],
\begin{equation}
 t=\frac{\pi}{2}\,\frac{1-\sigma}{\cos\left(\frac{\pi}{2}\sigma\right)}\,a(\tau),\quad
 r=\frac{\pi}{2}\sigma
\end{equation}
with
\begin{equation}
 a(\tau):=\begin{cases}
           \tau, & \textrm{problem (i)}\\
           \tau-1, & \textrm{problem (ii)}\\
           2\tau-1, & \textrm{problem (iii)}
          \end{cases}.
\end{equation}

Again we consider three different exact solutions and test how accurately we can reproduce them numerically. To this end, we make the following choices for the free function $F$ in the general solution \eqref{eq:gensol2} (or~\eqref{eq:gensol3}) as given in Table~\ref{tab:ExB}.

\begin{table}[ht]
 \begin{tabular}{|p{2.5cm}p{1.8cm}|}
  \hline
            & $F(x)$\\
  \hline
  Example 1: & $\cos(3x)$\\[1ex]
  Example 2: & $\frac{1}{10}x^3$\\[1ex]
  Example 3: & $\frac{1}{5}\sin(x^2)$\\[1ex]
  \hline  
 \end{tabular}
 \caption{\label{tab:ExB}
  Three example solutions to the conformally invariant wave equation that are regular everywhere in Minkowski space, specified in terms of the function $F(x)$, cf.~Eq.~\eqref{eq:gensol2} or \eqref{eq:gensol3}.}
\end{table}

First we consider the Cauchy problem (i). Here we prescribe the values of $f$ and $f_{,t}$ at $t=0$. Then we solve the wave equation \eqref{eq:WE2} with the following boundary conditions: at $r=0$ we impose $f_{,r}=0$, with exception of the point $r=t=0$, where we use the condition \eqref{eq:con2}; at $i^0$  we impose $f_{,t}=0$, but at the intersection of $i^0$ with $t=0$, where this condition is already satisfied by our choice of initial data, we have the condition $f_{,tt}=0$.

\begin{figure}\centering
 \includegraphics[width=\linewidth]{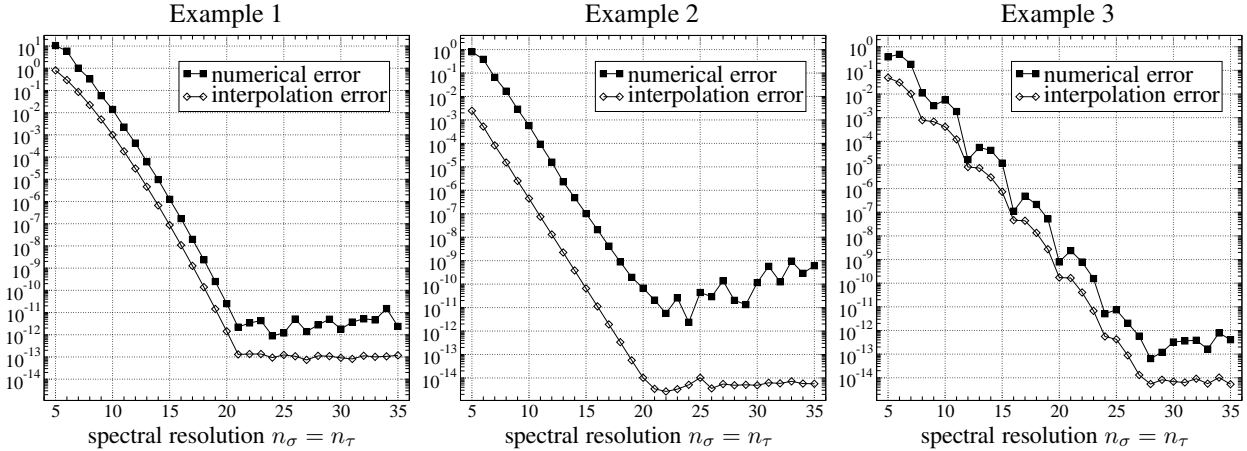}
 \caption{\label{fig:ConB}
  Convergence plots for the Cauchy problem from Fig.~\ref{fig:IVPs}b. The three examples are described in Table \ref{tab:ExB}.}
\end{figure}

Similarly to our numerical studies near $i^0$, we solve the problems with different resolutions and compare the numerical error to the interpolation error. The resulting convergence plots are shown in Fig.~\ref{fig:ConB}.

We observe a similar behaviour as for the wave equation near $i^0$: the convergence curves are initially approximately linear, corresponding to exponential convergence, until saturation near machine accuracy is reached. Moreover, the error due to the numerics is about 1-3 orders of magnitude larger than the interpolation error. In the case of example 2, we observe that the numerical error is again growing for $n_\sigma > 22$. This happens occasionally in the fully pseudospectral scheme due to accumulation of rounding errors for higher resolutions.

Next we consider the characteristic problems (ii) and (iii), where, in the latter case, we both solve the regular and singular wave equations.

For problem (ii), we prescribe the values of $f$ at $\Scri^-$. Then we solve the wave equation \eqref{eq:WE2} with the boundary condition $f_{,r}=0$ at $r=0$ and with condition \eqref{eq:con3} at the intersection of $r=0$ and $\Scri^-$.

In case of the characteristic problem (iii) for the regular wave equation, we again prescribe the values of $f$ at $\Scri^-$. However, this time this initial condition is not achieved with an expansion of the form \eqref{eq:expan} --- it turns out that this leads to numerical instabilities in the present case where the entire Minkowski space is considered. Instead, we just use the condition that $f$ be equal to the prescribed initial values at $\tau=0$, i.e.\ here the function values of $f$ themselves (and not $f_2$) are the unknowns in the numerical method. Furthermore, at $r=0$ we again impose the regularity condition $f_{,r}=0$.

Finally, in the characteristic problem (iii) for the singular wave equation \eqref{eq:WE3}, we prescribe values for $f_{,t}$ at $\Scri^-$. As in the regular case before, we impose this condition directly instead of enforcing it with an expansion. As boundary conditions we use $f_{,r}=0$ at $r=0$ and $f=0$ at $\Scri^+$ and $i^0$.

\begin{figure}\centering
 \includegraphics[width=\linewidth]{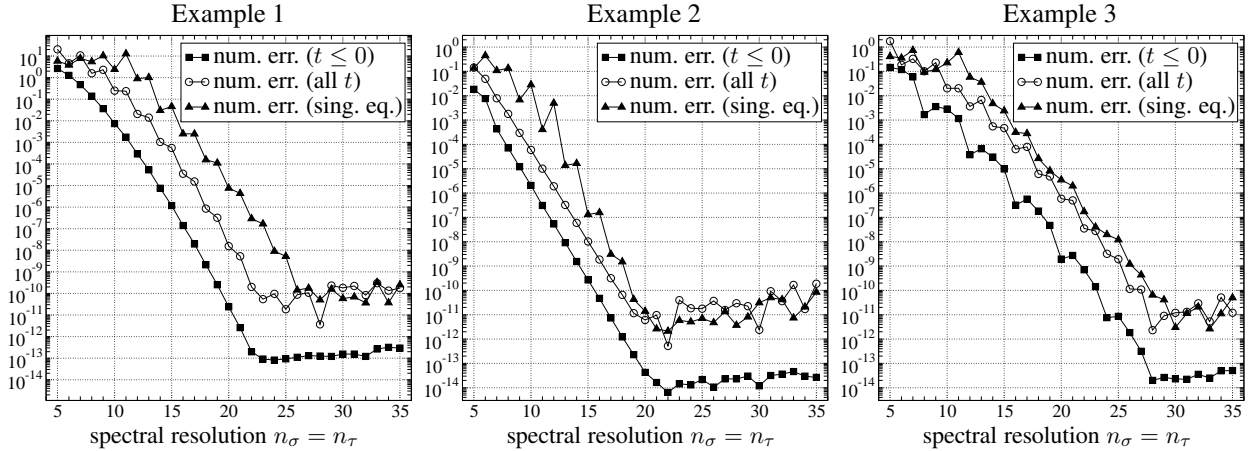}
 \caption{\label{fig:ConC}
  Convergence plots for the characteristic initial value problems. Again we have studied the three numerical examples described in Table \ref{tab:ExB}. Initial data are prescribed at $\Scri^-$ and then evolved until $t=0$ (see Fig.~\ref{fig:IVPs}c) or to $\Scri^+$ (Fig.~\ref{fig:IVPs}d). In the latter case, both the regular wave equation for the conformal metric and the singular wave equation corresponding to the physical metric are solved.}
\end{figure}

Convergence plots for the three described problems, using the same exact solutions from Table~\ref{tab:ExB} as before, are shown in Fig.~\ref{fig:ConC}. (This time we do not compare the numerical error with the interpolation error, which would lead to similar plots as above.)

As one should expect, we observe the smallest errors in the case where only the lower half $t\le0$ of Minkowski space is considered, whereas the accuracy drops if we evolve the data on $\Scri^-$ to $\Scri^+$, since the solution has then to be computed on a much larger domain. Interestingly, in the latter case, we obtain comparable errors\footnote{This is slightly different from the considerations in \cite{Hennig2009}, where a hyperboloidal initial value problem was considered. In an example studied in this paper, the singular equation could be solved with even more accuracy than the regular equation. We also note that, if we were to use the numerically computed function $\tilde f$ to calculate the conformally related function $f=\tilde f/\Theta$, then we would loose some accuracy. For, applying L'H\^opital's rule, we would need to differentiate $\tilde f$ at the boundaries (where $\Theta=0$). And with each differentiation of a function approximated in terms of Chebyshev polynomials, the error increases by 
about 1 order of magnitude.} for the regular and singular formulations --- the only difference being that for the singular case, slightly higher resolutions are needed to reach the saturation level. This indicates that, from a numerical point of view, the regular formulation of an equation is not necessarily preferred to a singular formulation --- provided suitable numerical techniques are applied to treat the singular case.

\begin{figure}\centering
 \includegraphics[width=0.6\linewidth]{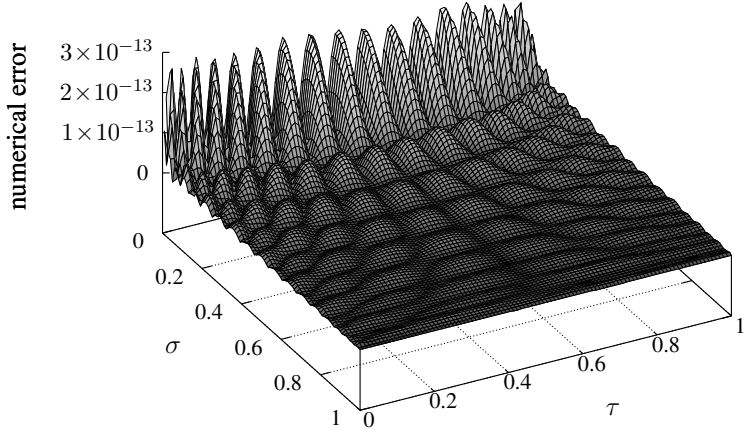}
 \caption{\label{fig:ErrB}
 The numerical error for the (regular) wave equation in the entire Minkowski space. The plot is obtained for example solution 2 from Table~\ref{tab:ExB} with spectral resolution $22\times 22$.}
\end{figure}

An example for the distribution of the error over the numerical domain is shown in Fig.~\ref{fig:ErrB} (which should be compared to the similar plot in Fig.~\ref{fig:ErrA}). This time the largest error is assumed at or near $\sigma=0$, corresponding to the origin $r=0$ of Minkowski space. Also note that the error is not zero at $\tau=0$, since for this calculation the initial conditions are not enforced exactly, as described above. In some numerical examples, a peak in the error is found at $\sigma=0$, $\tau=1$, but otherwise we obtain very similar plots for other examples and resolutions.

%%%%%%%%%%%%%%%%%%%%%%%%%%%%%%%%%%%%%%%%%%%%%%%%%%%%%%%%%%%%%%%%%%%%%%%%%
\section{Discussion\label{sec:discuss}}

As a model problem for solving Einstein's equations near the cylinder at spacelike infinity $i^0$, we have considered the conformally invariant wave equation on Minkowski background. With the fully pseudospectral scheme from \cite{Hennig2009}, it was possible to obtain highly accurate numerical solutions (with numerical error in the order $10^{-10}$ to $10^{-13}$, close to machine accuracy of our double precision code) for a small number of grid points. In particular, the method does not fail at $\Scri$ or at the intersection $I^\pm$ of the cylinder and $\Scri^\pm$, where the equation degenerates. Instead, it was possible to include both $\Scri$ and $I^\pm$ into the numerical domain. This is in contrast to finite-difference approaches, where $\Scri$ can only be reached asymptotically as a consequence of the CFL condition. 

We have also compared the numerical treatment of the regular wave equation (with respect to the conformal metric) to the singular wave equation (with respect to the ``physical'' metric). In both cases we have achieved comparable numerical accuracy, which shows that both formulations have equal rights from a numerical point of view (whereas the ``regular formulation'' is preferential in analytical approaches, since formally singular terms complicate the calculations).

These observations suggest that our fully pseudospectral scheme should be a promising method also for treating more general equations (e.g., the spin-2 system or the full Einstein equations) and less symmetric spacetimes (e.g., axisymmetric\footnote{It was recently demonstrated for the first time that fully pseudospectral time evolution can be applied to $2+1$ dimensional problems \cite{Macedo2013}.} or even completely non-symmetric solutions).

%%%%%%%%%%%%%%%%%%%%%%%%%%%%%%%%%%%%%%%%%%%%%%%%%%%%%%%%%%%%%%%%%%%%%%%%%
\begin{acknowledgments}
 We would like to thank Robert Thompson for commenting on the manuscript.
 This work was supported by the Marsden Fund Council from Government funding, administered by the Royal Society of New Zealand.
\end{acknowledgments}

%%%%%%%%%%%%%%%%%%%%%%%%%%%%%%%%%%%%%%%%%%%%%%%%%%%%%%%%%%%%%%%%%%%%%%%%%

%%%%%%%%%%%%%%%%%%%%%%%%%%%%%%%%%%%%%%%%%%%%%%%%%%%%%%%%%%%%%%%%%%%%%%%%%

\begin{thebibliography}{99}

\bibitem{Ansorg2002}
Ansorg, M., Kleinw\"achter, K., and Meinel, R.,
 \emph{Highly accurate calculation of rotating neutron stars},
 Astron.\ Astrophys.\ {\bf 381}, L49 (2002) 

\bibitem{Ansorg2003}
Ansorg, M., Kleinw\"achter, K., and Meinel, R.,
 \emph{Highly accurate calculation of rotating neutron stars: Detailed description of the numerical methods},
 Astron.\ Astrophys.\ {\bf 405}, 711 (2003)

\bibitem{Ansorg2011}
 Ansorg, M.\ and Hennig, J.,
 \emph{The interior of axisymmetric and stationary black holes: numerical and analytical studies},
 J.\ Phys.: Conf.\ Ser. {\bf 314}, 012017 (2011)

\bibitem{BeyerDoulis2012}
 Beyer, F., Doulis, G., Frauendiener, J., Whale, B., 
 \emph{Numerical space-times near space-like and null infinity. The spin-2 system on Minkowski space}, 
 Class.\ Quantum Grav.\ {\bf 29}, 245013 (2012)

\bibitem{BeyerDoulis2013}
 Beyer, F., Doulis, G., Frauendiener, J., Whale, B., 
 \emph{The spin-2 equation on Minkowski background}, 
 Springer Proc.\ Math.\ Stat.\ {\bf 60}, 465 (2014) 

\bibitem{Chrusciel2013}
 Chru\' sciel, P.~T., Paetz, T.-T., 
 \emph{Solutions of the vacuum Einstein equations with initial data on past null infinity}, 
 Class.\ Quantum Grav.\ {\bf 30}, 235037 (2013)


\bibitem{CruzOsorio2010}
 Cruz-Osorio, A., Gonz\'alez-Ju\'arez, A., Guzm\'an, F.~S., Lora-Clavijo, F.~D.,
 \emph{Numerical solution of the wave equation on particular space-times using CMC slices and scri-fixing conformal compactification},
 Rev.\ Mex.\ F\' is.\ {\bf 56}, 456 (2010)

 \bibitem{Kannar1996}
 K\'ann\'ar, J., 
 \emph{On the existence of $C^\infty$ solutions to the asymptotic characteristic initial value problem in general relativity}, 
 Proc.\ Roy.\ Soc.\ London A {\bf 452}, 945 (1996)

\bibitem{Doulis2013}
 Doulis, G., Frauendiener, J., 
 \emph{The second order spin-2 system in flat space near space-like and null-infinity}, 
 Gen.\ Relativ.\ Gravit.\ {\bf 45}, 1365 (2013)

\bibitem{Friedrich1981} 
 Friedrich, H., 
 \emph{On the regular and the asymptotic characteristic initial value problem for Einstein's vacuum field equations},
 Proc.\ Roy.\ Soc.\ London A {\bf 375}, 169 (1981)

\bibitem{Friedrich1983}
 Friedrich, H.,
 \emph{Cauchy problems for the conformal vacuum field equations in general relativity},
 Commun.\ Math.\ Phys.\ {\bf 91}, 445 (1983)

\bibitem{Friedrich1986a}
 Friedrich, H., 
 \emph{On purely radiative space-times}, 
 Comm.\ Math.\ Phys.\ {\bf 103}, 35 (1986)

\bibitem{Friedrich1986}
 Friedrich, H.,
 \emph{On the existence of $n$-geodesically complete or future complete solutions of Einstein's field equations with smooth asymptotic structure},
 Commun.\ Math.\ Phys.\ {\bf 107}, 587 (1986)

\bibitem{Friedrich1998}
 Friedrich, H.,
 \emph{Gravitational fields near space-like and null infinity},
 J.\ Geom.\ Phys.\ {\bf 24}, 83 (1998)

\bibitem{Friedrich2003}
 Friedrich, H.,
 \emph{Spin-2 fields on Minkowski space near spacelike and null infinity},
 Class.\ Quantum Grav.\ {\bf 20}, 101 (2003)

\bibitem{Friedrich2013}
 Friedrich, H.,
 \emph{The Taylor expansion at past time-like infinity},
 Commun.\ Math.\ Phys.\ {\bf 324}, 263 (2013)

\bibitem{Frauendiener2004}
 Frauendiener, J.,
 \emph{Conformal infinity},
 Living Rev.\ Relativity {\bf 7},  (2004)

\bibitem{Gundlach1994}
 Gundlach, C., Price, R.\ H. and Pullin, J.,
 \emph{Late-time behavior of stellar collapse and explosions. I. Linearized perturbations},
 Phys.\ Rev.\ D {\bf 49}, 883 (1994)

\bibitem{Hennig2009}
 Hennig, J.\ and Ansorg, M.,
 \emph{A fully pseudospectral scheme for solving singular hyperbolic equations on conformally compactified space-times},
 Journal of Hyperbolic Differential Equations {\bf 6}, 161 (2009)

\bibitem{Hennig2013}
 Hennig, J.,
 \emph{Fully pseudospectral time evolution and its application to $1+1$ dimensional physical problems},
 J.\ Comput.\ Phys.\ {\bf 235}, 322 (2013)

\bibitem{Macedo2013}
 Panosso Macedo, R.\ and Ansorg, M.,
 \emph{Fully spectral code for linear axisymmetric wave equations on hyperboloidal foliations},
 Talk given at the GR20 conference, Warsaw (2013)

\bibitem{Penrose1964}
 Penrose, R.,
 \emph{The light cone at infinity}, in \emph{Relativistic theories of gravitation},
 ed Infeld, L., Pergamon Press, Oxford (1964)

\bibitem{Penrose1965}
 Penrose, R.,
 \emph{Zero rest-mass fields including gravitation: asymptotic behaviour},
 Proc.\ Roy.\ Soc.\ London A {\bf 284}, 159 (1965) 
 
\bibitem{Zenginoglu2008}
 Zengino\u glu, A., 
 \emph{A hyperboloidal study of tail decay rates for scalar and Yang-Mills fields},  Class.\ Quantum Grav.\ {\bf 25}, 175013 (2008)

\end{thebibliography}
\end{document}